\documentclass[twocolumn,showpacs,preprintnumbers,amsmath,amssymb]{revtex4}

\usepackage{graphicx}
\usepackage{dcolumn}
\usepackage{bm}

\begin{document}

\preprint{}

\title{Holes in the valence band of superconducting boron-doped diamond film \\
studied by soft X-ray absorption and emission spectroscopy}

\author{Jin Nakamura,$^1$\email, Tamio Oguchi,$^2$  
Nobuyoshi Yamada,$^1$  Kazuhiko Kuroki,$^1$ Kozo Okada,$^3$ Yoshihiko Takano,$^4$ 
Masanori Nagao,$^4$ Isao Sakaguchi,$^4$ Hiroshi Kawarada,$^5$
Rupert C.C. Perera,$^6$ and David L. Ederer$^7$}%

\affiliation{%
$^1$Department of Applied Physics \& Chemistry, The University of
Electro-Communications,
Chofu, Tokyo 182-8585, Japan
}%
\affiliation{
$^2$Department of Quantum Matter, ADSM, Hiroshima University,
Higashihiroshima, Hiroshima 739-8530, Japan
}%
\affiliation{
$^3$Department of Physics, Okayama University,
Tsushima-naka, Okayama 700-8530, Japan 
}%
\affiliation{
$^4$National Institute for Materials Science,
Tsukuba, Ibaraki 305-0047 Japan
}%
\affiliation{
$^5$School of Science and Engineering, Waseda University,
Shinjyuku-ku, Tokyo, 169-8555, Japan
}%
\affiliation{
$^6$Center for X-ray Optics, Lawrence Berkeley National Laboratory,
Berkeley, CA 94720, USA
}%
\affiliation{
$^7$Department of Physics, Tulane University,
New Orleans, Louisiana 70118, USA
}%
\date{\today}

\begin{abstract}
Carbon- and boron-2$p$ states of 
superconducting and non-superconducting 
boron-doped diamond samples are measured using soft X-ray
emission and absorption spectroscopy.  
For the superconducting sample, a large density of hole
states is observed in the valence band in addition to the states in the 
impurity band. 
The hole states in the valence band is located at about 1.3 eV below 
the valence band maximum regardless of the doping level, which 
cannot be interpreted within a simple rigid band model.
Present experimental results, combined with the first principles 
calculations, suggest that superconductivity is to be attributed to the holes
in the valence band.
\end{abstract}
\maketitle

Recent discovery of superconductivity in heavily boron-doped
diamond\cite{ekimov} has brought up renewed interest in the physics of doped 
insulating/semiconducting systems.
Diamond has always been a very attractive material because of its characteristic
physical properties, and 
boron has been well known as an impurity in diamond along with 
nitrogen and phosphorus, 
which makes an acceptor level at 0.37 eV above the top of the valence band.\cite{thonke}  
Heavily boron-doped diamond has been known to be metallic since the late '90s  
(the concentration for insulator-metal transition $c_{\rm M}$ =
2$\times$10$^{20}$ cm$^{-3}$), but 
due to the large band gap ($\sim$ 5.5 eV) the possibility
of superconductivity in diamond had not been noticed for some time, 
despite the fact that graphite-related compounds and diborides have been studied
extensively.\cite{nagamatsu}

Since the discovery of superconductivity in boron-doped diamond,
several theoretical proposals have been made.
\cite{boeri,lee,xiang,blase,baskaran}
Theories based on conventional BCS theory are presented in an
analogy with MgB$_2$, where strong coupling
of the holes at the top of the valence band to the
optical phonons plays an important role.\cite{boeri,lee,xiang,blase}
Boeri\cite{boeri} and Lee\cite{lee} argued the superconductivity in the
hole states of valence band by virtual crystal approximation, assuming that the
crystal is made "averaged carbon-bore atoms", 
and pointed out the importance of the strong electron-phonon coupling.
Xiang\cite{xiang} and Blase\cite{blase} discussed the effect of boron impurity 
on the superconductivity using density functional calculations on the system
including boron as impurity.
Secondly, Baskaran pointed out the importance of the electron correlation
in the boron-impurity band based on the resonating valence
band (RVB) theory.\cite{baskaran}

Now, under these circumstances, 
an important step towards understanding the mechanism of 
superconductivity is to identify the character of the carriers.  
Then, it is a prerequisite to perform an experimental study which 
determines whether the holes 
responsible for superconductivity reside in the impurity band or valence band.
Unfortunately, there are  few experimental works at present, 
which may be due to experimental difficulties in observing superconductivity.
Recently, however, Takano et al. have successfully synthesized a
superconducting diamond film using a conventional microwave plasma-assisted
chemical vapor deposition (MPCVD) method.\cite{takano}

Soft X-ray absorption (XAS) and emission (XES) spectroscopies near C-$K$
and B-$K$ edges are powerful tools for studying the electronic structure of the 
boron-doped diamond.
As the dipole selection rule between the 1$s$ core level and the 2$p$ state 
governs the transition, we can obtain the partial density of 
states (PDOS) of carbon 2$p$ and of boron 2$p$ separately, 
which can give important information on the mechanism of 
superconductivity.\cite{jin_mgb2p,jin_alb2s,jin_mgb2s}
Moreover, XES spectroscopy is a bulk sensitive probe, not a surface 
sensitive one like photoelectron spectroscopy.

In a previous publication we paid attention to the electronic structure of
the boron impurities in diamond, a non-superconducting sample, 
and performed XAS and XES measurements near the B-$K$ and the C-$K$ edges.\cite{jin_dia}
Our conclusion there is that the wave function of impurity B-2$p$ strongly
hybridizes with the host C-2$p$ in lightly boron-doped diamond specimen.

In this Letter, we study the electronic structure of both B-2$p$ and
C-2$p$ in the superconducting and non-superconducting 
diamond films using XAS and XES spectroscopy
near the B-$K$ and the C-$K$ edges. 
The experimental results are further analyzed in detail 
with the aid of first-principles band calculations.
For the superconducting sample, a large electronic density of hole states
is observed in the valence band in addition to the states in the impurity band.  
The hole states in the valence band is located at about 1.3 eV below the
valence band maximum, which is larger than what is expected 
from the band calculation  results. Moreover, this energy is 
not sensitive to the doping level, which 
cannot be interpreted within a simple rigid band picture. 
Although the impurities do form a band, 
superconductivity is attributed to the holes in valence band
because the Fermi level lies within this band.

Boron-doped diamond thin films were deposited on Si (001) wafers in
MPCVD.\cite{takano,kamo,yano}
Three diamond samples with different boron-concentrations were used;
sample labeled "sample-S"shows superconductivity at $T_{\rm c}$ = 3.7 K,
and labeled "sample-N1" and "-N2" do not show $T_{\rm c}$ above 1.75 K
using SQUID measurements.
The boron concentrations are estimated by SIMS measurement to be about 
4.3at\%B,
0.25at\%B and 0.10at\%B for the sample-S, -N1 and N2, respectively.

The XAS and XES measurements were performed at BL-8.0.1\cite{alsbl8}
of the Advanced Light Source (ALS) in Lawrence Berkeley National Laboratory
(LBNL).
The energy resolutions of the incoming and outgoing X-rays were 0.2 and 0.4
eV, respectively.
For the calibrations of the monochromator and spectrometer, $h$-BN, HOPG and
natural diamond were used as the standard samples.\cite{ma,skytt,muramatsu}

\begin{figure}
\includegraphics[width=7cm]{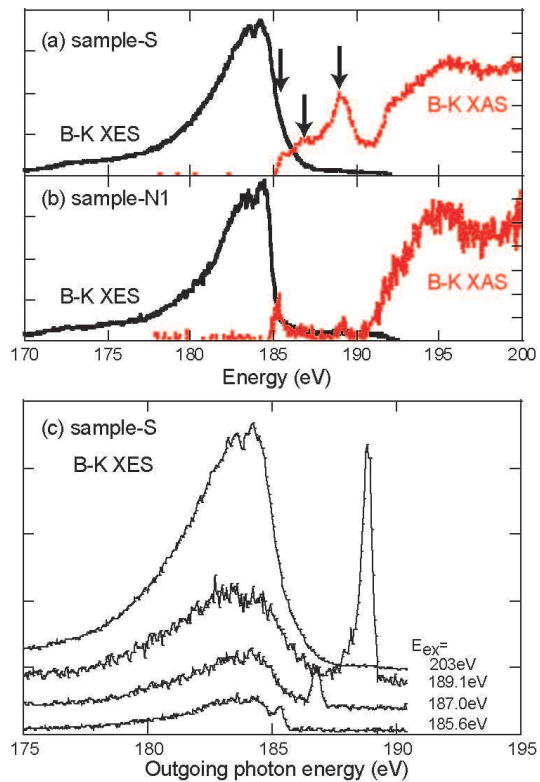}
\caption{\label{Fig1}
B-$K$ XAS and XES spectra of (a) sample-S and
(b)sample-N1.
The incident angle was set to $\theta$=20$^{\circ}$.
The excitation energy of XES measurement was 200 eV.
The area intensity of XES is normalized to unity.
(c) Several B-$K$ XES spectra of sample-S.
The area intensity of XES is normalized to the peak height at the
excitation energy in XAS spectra indicated by the arrows in (a).
}
\end{figure}

As mentioned in the introductory part, the B-$K$ XAS and XES spectra represent 
PDOS of empty and occupied B-2$p$ states of the {\it impurity boron}, respectively.
Figure~\ref{Fig1} shows B-$K$ XAS and XES spectra of sample-S and -N1.
In this figure [Fig.~\ref{Fig1} (a) and (b)], the excitation energy of the 
XES spectra was 200 eV in both cases.
One can see the good agreement between the XES spectra of the two samples, but 
in the XAS spectra there is a remarkable difference in the host-gap region,
185 $\sim$ 190 eV.
The B-$K$ XAS of sample-S shows in-gap states, i.e.,
a broad state which fills the whole band gap region and a sharp peak
at about 189 eV.
These in-gap states may be attributed to boron at interstitial sites 
and/or boron forming clusters.
The B-$K$ XAS of sample-N1 is essentially the same as our previous 
result of a non-superconducting sample\cite{jin_dia} except a small trace of the peak
at 189 eV.
The Fermi level is estimated to be located at about 185.1 eV in both samples,
which is a little bit lower than observed in the previous report, 185.3 eV.
We attribute the difference to experimental error.

Figure~\ref{Fig1}(c) shows the B-$K$ XES with several excitation energies of 
sample-S, where three of them have the excitation energy 
of the in-gap states and one of them is a normal XES
spectra with high energy excitation ($E_{\rm ex}$= 203 eV).
For all the XES spectra of sample-S, the overall features are almost the same with
each other and also the same with the XES of sample-N1, which is consistent with
the previous report.\cite{jin_dia}

\begin{figure}
\includegraphics[width=7cm]{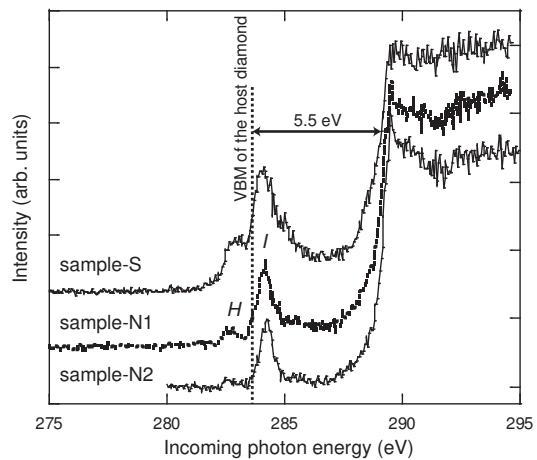}
\caption{\label{Fig2}
The C-$K$ XAS spectra of sample-S and sample-N1.
The spectrum of the previous report is also shown as sample-N2.
The incident angle was set to $\theta$=70$^{\circ}$.
The dotted line represents the energy of the
valence band maximum using the band gap of 5.5 eV.
Two characteristic peaks, $H$ and $I$, are observed at about
282.6 and 284.1eV.  }
\end{figure}

Figure~\ref{Fig2} shows the C-$K$ XAS spectra of these samples.
There is no incident angle dependence in the XAS and XES (following) spectra,
which suggests that there is no effect of preferred orientation
of nano-size crystals.
The XAS spectra of sample-N1 and -N2 are almost identical,  
showing the impurity in-gap state labeled $I$ at 284.1 eV. However, except a small
state ($H$) at 282.5 eV.
In the superconductor, sample-S, this state density ($H$) increases 
considerably.
This state is located at about 1.3 eV below the valence band maximum (VBM) of the
non-doped diamond at 283.8 eV, which is shown by a dotted line.\cite{jin_dia}
Therefore, in the diamond superconductor, there is a considerable amount of holes
in the C-2$p$ valence band.  
\begin{figure}
\includegraphics[width=7cm]{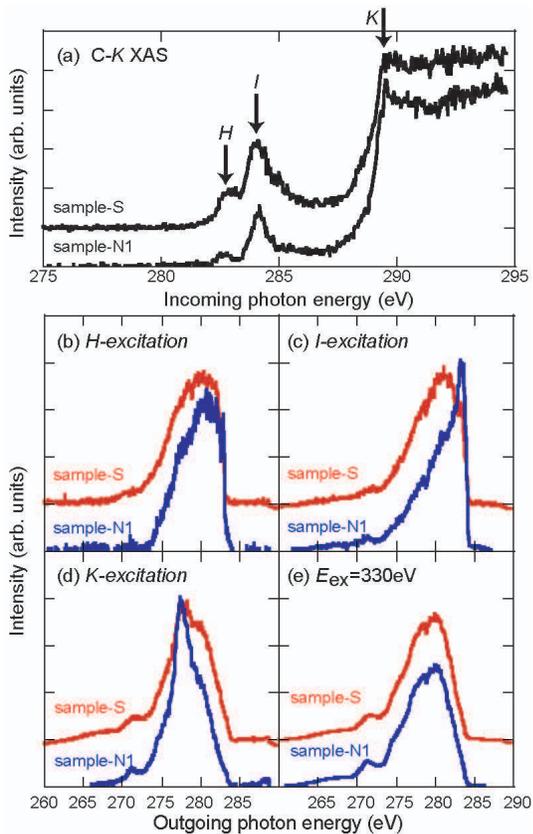}
\caption{\label{Fig3}
(a) C-$K$ XAS and (b-e) C-$K$ XES spectra of superconductor (sample-S) and 
non-superconductor
(sample-N1).
The area intensities of all XES spectra are normalized to unity.
The excitation energy is presented in XAS spectrum (a) as an arrow.  }
\end{figure}

In order to gain more insight into the electronic states involved with the
hole states, we performed XES measurements with an excitation energy of 282.6 eV
(corresponding to state-$H$ in Fig.~\ref{Fig2}), 284.1 eV (corresponding to $I$), 
and the energy corresponding to the 
bottom of the conduction band ($K$). 
Figure~\ref{Fig3} shows the C-$K$ XES spectra of sample-S and -N1,
where $H$-, $I$- and {\it K-excitation} denotes the corresponding excitation energy.
At the excitation energy near the Fermi level, namely the {\it H-excitation} 
[Fig.~\ref{Fig3}(b)], the XES spectra of both  sample-S and -N1 are the same
and similar to the XES of the host diamond.
This suggests that state-$H$ in the XAS spectra [Fig.~\ref{Fig3}(a)]
is a state of bulk diamond, not a state of precipitation.
In the {\it I-excitation} XES [Fig.~\ref{Fig3}(c)], one can see the difference
between sample-S and -N1, i.e.,
the XES of sample-S is similar to the one with {\it H-excitation},
while that of sample-N1 displays a sharp elastic peak and a large tail on the lower
energy side.
The appearance of the sharp elastic peak and the large tail 
agrees with our previous results of non-superconductors.
Since sample-S and sample-N1 differs in the density of impurities, 
the fact that the I-excitation XES spectra differs between the two samples 
suggest that  state-$I$ is essentially an impurity state.  
In the case of {\it K-excitation}, a small difference is observed between the two samples.
The spectrum of sample-N1 shows a sharp peak at 277.4 eV, which corresponds to
doubly degenerated states of the C-2$p$ valence band at $X_4$ point
because the bottom of the conduction band (C.B.) is at $X_1$ point in 
diamond.\cite{ma2}
This sharp peak broadens in sample-S, and moreover, the sharp XAS-edge at 289.6 eV
(the bottom of C.B.) broadens in sample-S (Fig.~\ref{Fig2}).  
This may be because the symmetry of the system becomes lower due to 
atom substitution and/or possible lattice distortion upon boron doping.  
The normal XES spectra are in good agreement with each other, 
and with XES of the host-diamond.

\begin{figure}
\includegraphics[width=7cm]{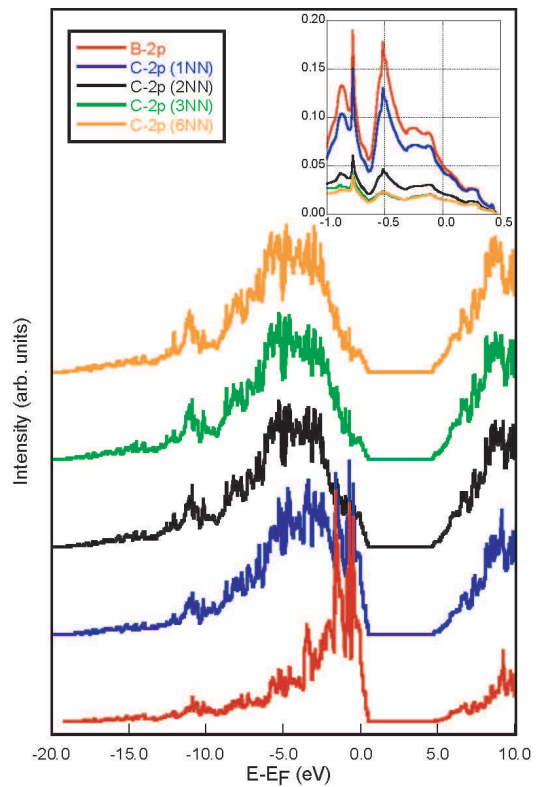}
\caption{\label{Fig4}
The energy dependence of the PDOS 
of B- and C-2$p$ states obtained by  first-principles band calculations
using a super-cell (nominally C$_{63}$B$_1$).
$n$NN indicates the $n$-th nearest neighbor carbon from the
impurity boron atom.
Among the six types of carbons, only 1NN, 2NN, 3NN and 6NN C are shown in the
figure.
The inset shows the PDOS around the Fermi level.
}
\end{figure}
In order to analyze the experimental spectra, we performed first-principles 
band-structure
calculations for a diamond with substitutional single B impurity using two 
kinds of
super-cell [nominally C$_{31}$B$_1$ (3.1at\%B) and C$_{63}$B$_1$ (1.6at\%B)].
Figure~\ref{Fig4} shows the calculated 2$p$ PDOS
using the C$_{63}$B$_1$ super-cell.
1NN, 2NN, 3NN and 6NN C indicate 1st, 2nd, 3rd and 6th nearest neighbor carbon
from the impurity boron atom, respectively.
The density of states of the first nearest neighbor carbon (1NN C) is large
at around the Fermi energy and has hole states above the Fermi energy. This
feature is consistent with the result obtained by DV-X$\alpha$
method in our previous study.\cite{jin_dia}
In contrast with this, the density of states of 6NN C, which is far from
the impurity boron, is similar to the PDOS of the non-doped diamond (bulk), but
shows a small number of holes around the Fermi level (see the inset of 
Fig.~\ref{Fig4}).
For the C$_{31}$B$_1$ super-cell, heavy-doped case, the density of states 
of 4th neighbor carbon
(4NN C, not shown) which is farthest from the impurity boron in this cell has
a notable number of holes above the Fermi energy,
i.e., the hole state around the Fermi level
increases with an increase of boron concentration.

Now let us go back to the data for the C-$K$ edge with these band calculation 
results in mind. The XES result at energy-$I$
excitation of sample-N1 [Fig.~\ref{Fig3}(b)] is very similar to the PDOS of 
first nearest neighbor carbons (Fig.~\ref{Fig4}).
Therefore, the impurity states can be interpreted to be of localized one in the 
non-superconducting sample.
However, for sample-S, the XES at energy-$I$ excitation is very similar to
the PDOS of the bulk diamond.
This means the carbons involving the impurity hole states are of bulk carbons, 
i.e. the impurity states make a band.
Thus, we may conclude that the impurity state of sample-N1
is localized while the impurity state of sample-S forms a band.

The XES results at excitation energy-$H$ of sample-S and -N1, shown
in Fig.~\ref{Fig3}(b), are very similar to the PDOS of bulk carbons rather 
than that of first nearest neighbor carbon.
As energy $H$ is located at about 1.3 eV below VBM,
we attribute the hole states at energy $H$ to the hole states in the 
valence band.  
Since this valence band hole state lies right next to the Fermi level,
we may conclude that the holes in the valence band plays an essential 
role in the occurrence of superconductivity.  

If we look more quantitatively into the energy of the peak $H$, 
its location seems to be somewhat lower than what is expected from 
the first-principles calculation. Namely, the Fermi energy obtained 
from our band calculation 
is 0.7 eV and 0.5 eV below the VBM for 32-atom and  64-atom super-cells, 
respectively, which is smaller than 1.3 eV.
More interestingly, the energy of peak $H$ does not depend on the 
doping level, and only its XAS intensity grows with doping (see Fig.~\ref{Fig2}), 
which cannot be understood within a simple rigid band picture.
This curious doping dependence of peak $H$ remains 
as an interesting future problem.

In conclusion, the partial density of states of the C-2$p$ state and the B-2$p$
states of superconducting and non-superconducting boron-doped diamond samples have 
been measured using soft 
X-ray emission and absorption spectroscopy on C-$K$ and B-$K$ edges.
For the superconducting sample, a large electronic density of states of holes has been 
observed in the valence band in addition to the states in impurity band.
However, the hole states in valence band located at around 1.3 eV below the valence
band maximum cannot be interpreted within a simple rigid band model.
Although our results show that 
the impurity states do form a band in the superconducting sample,
the holes in the valence band seems to be more important for superconductivity 
because the Fermi level lies within the valence band.

We express our thanks to Prof. M. Tachiki of National Institute for 
Materials Science (NIMS)
and Prof. A. Natori of University of Electro-Communications for useful 
discussions.
This work was performed under the approval of ALS-LBNL, proposal No. 
ALS-00931.
ALS is supported by the Director, Office of Science, Office of Basic Energy 
Sciences,
Materials Sciences Division, of the U.S. Department of Energy
under Contract No. DE-AC03-76SF00098 at LBNL.
%


\end{document}